\documentstyle[preprint,prd,aps]{revtex}
\begin{document}
\draft
\begin{titlepage}
\preprint{\vbox{\hbox{UDHEP-06-98}
\hbox{June 1998} }}
\title{Chiral Symmetry Breaking in an External Field\footnote{Talk 
presented at the 2nd Latin American Symposium on 
High Energy Physics, San Juan, Puerto Rico, April 8-11, 1998.}} 
\author{C. N. Leung}
\address{Department of Physics and Astronomy, 
University of Delaware\\
Newark, DE 19716 \\}
\maketitle

\begin{abstract}

The effects of an external field on the dynamics of chiral 
symmetry breaking are studied using quenched, ladder QED as our 
model gauge field theory.  It is found that a uniform external 
magnetic field enables the chiral symmetry to be spontaneously 
broken at weak gauge couplings, in contrast with the situation 
when no external field is present.  The broken chiral symmetry 
is restored at high tempeatures as well as at high chemical 
potentials.  The nature of the two chiral phase transitions is 
different: the transition at high temperatures is a continuous 
one whereas the phase transition at high chemical potentials is 
discontinuous.

\end{abstract}
\end{titlepage}

I would like to report here some recent 
work\cite{LNA,LLN} done in collaboration primarily 
with D.-S. Lee and Y. J. Ng.  The objective is to 
understand how an external field may affect the 
dynamics of chiral symmetry breaking in gauge theories.  
Quantum electrodynamics (in $3+1$ dimensions), treated 
in the quenched, ladder (or planar) approximation, will 
serve as our model gauge theory.  At present our 
formalism is applicable only for a constant (in space 
as well as in time) external field and I shall discuss 
results pertinent to the case of a constant magnetic field.  

The problem of dynamical chiral symmetry breaking in a 
uniform magnetic field in quenched, planar QED has 
received some attention in recent years\cite{LNA} 
-\cite{Mir}.  Following the pioneering works\cite{csb} 
on the study of chiral symmetry breaking in quenched, 
ladder QED, our approach is based on the Schwinger-Dyson 
(SD) equation for the fermion self-energy.  In this 
approach, one starts with a Lagrangian with a zero bare 
fermion mass and looks for nontrivial solutions to the 
SD equation which signal a dynamically generated 
fermion mass and a possible spontaneous breakdown of the 
chiral symmetry.

Let us consider QED with a single charged fermion having 
a zero bare mass and charge $g$.  In the quenched, ladder 
approximation, the SD equation in the $x$-representation 
reads
\begin{equation}
M(x,x') = i g^2 \gamma^\mu G_A(x,x') \gamma^\nu D_{\mu\nu}
(x-x'),
\label{SDx}
\end{equation}
where $M(x,x')$ is the fermion mass operator in the 
$x$-representation, $M(x,x') = \langle x| \hat{M} |x' 
\rangle$, $D_{\mu\nu}(x-x')$ is the bare photon propagator,
and $G_A(x,x')$ is the fermion propagator in the presence of 
an external field represented by the vector potential 
$A_\mu (x)$.  We adopt the metric with the signature 
$g_{\mu \mu} = (-1, 1, 1, 1)$.  $G_A$ satisfies the equation,
\begin{equation}
\gamma^\mu \Pi_\mu G_A(x,y) + \int d^4x' M(x,x') G_A(x',y) 
= \delta^{(4)}(x-y),
\label{Greeneq}
\end{equation}
where $\Pi_\mu \equiv - i \partial_\mu - g A_\mu (x)$.  
Schwinger\cite{Sch} was the first to obtain an exact 
analytical expression for the fermion Green's function in 
the presence of a constant electromagnetic field of 
arbitrary strength.  However, we find the alternative 
representation of $G_A(x,y)$ proposed by Ritus\cite{Ritus} 
more convenient for our purpose.  The essence of this 
approach is explained below.

Since, in the presence of a constant external field, the 
fermion asymptotic states are no longer free particle 
states represented by plane waves, but are described by 
wavefunctions consistent with the particular external 
field configuration, namely, eigenfunctions of 
$(\gamma^\mu \Pi_\mu)^2$:
\begin{equation}
- (\gamma \cdot \Pi)^2  \psi_p(x) = p^2 \psi_p(x).
\label{eigeneq}
\end{equation}
Instead of the usual momentum space, it is more convenient 
to work in the representation spanned by these 
eigenfunctions.  Another advantage of using this 
representation is that, for constant external fields, 
the mass operator is diagonal\cite{Ritus}.

If we work in the chiral representation of the Dirac 
matrices in which $\gamma_5$ and $\Sigma_3 = i \gamma_1 
\gamma_2$ are both diagonal with eigenvalues 
$\chi = \pm 1$ and $\sigma = \pm 1$, respectively, the 
eigenfunctions $\psi_p(x)$ has the general form
\begin{equation}
\psi_p(x) = E_{p\sigma\chi}(x) \omega_{\sigma\chi},
\label{eigenfcn0}
\end{equation}
where $\omega_{\sigma\chi}$ are bispinors which are the 
simultaneous eigenvectors of $\Sigma_3$ and $\gamma_5$.  
The exact functional form of the $E_{p\sigma\chi}(x)$ 
will depend on the specific external field configuration. 

In the case of a constant magnetic field of strength $H$ 
pointing in the $z$-direction, the vector potential may be 
taken to be $A_\mu = (0, 0, Hx_1, 0)$ and one finds that 
the eigenfunctions $E_{p\sigma\chi}(x)$ do not depend on 
$\chi$:
\begin{equation}
E_{p\sigma}(x) = N {\rm e}^{i (p_0x^0 + p_2x^2 + p_3x^3)} 
D_n(\rho). 
\label{eigenfcn}
\end{equation}
Here $N$ is a normalization factor and $D_n(\rho)$ are the 
parabolic cylinder functions\cite{math} with argument 
$\rho \equiv \sqrt{2 |g H|} (x_1 - \frac{p_2}{g H})$ and  
index $n$ which labels the Landau levels:
\begin{equation}
n = n(k,\sigma) \equiv k + \frac{g H \sigma}{2 |g H|} - 
\frac{1}{2},~~~~n = 0, 1, 2, ...
\label{index}
\end{equation}         
The eigenvalue $p$ now stands for the four quantum numbers 
$(p_0, p_2, p_3, k)$, where $k$ is the discrete quantum 
number of the quantized squared transverse momentum:
\begin{eqnarray}
- (\gamma \cdot \Pi_\perp)^2  \psi_p(x) && ~\equiv~ 
- (\gamma^1 \Pi_1 + \gamma^2 \Pi_2)^2 \psi_p(x) 
\nonumber\\ && ~=~ 2 |gH| k \psi_p(x).
\label{pperp}
\end{eqnarray}
For a given $n$, the allowed values for $k$ are $k = n, n+1$.

The eigenfunction-matrices $E_p(x)$ defined as  
\begin{eqnarray}
E_p(x) ~&&~\equiv~\sum_\sigma E_{p\sigma}(x) {\rm diag} 
(\delta_{\sigma 1}, \delta_{\sigma -1}, \delta_{\sigma 1}, 
\delta_{\sigma -1}) \nonumber\\ ~&&~\equiv~ 
\sum_\sigma E_{p\sigma}(x) \Delta(\sigma)
\label{ep}
\end{eqnarray}
satisfy the orthonormality and completeness relations 
$(\bar{E}_p \equiv \gamma^0 E_p^\dagger \gamma^0)$:
\begin{eqnarray}
\int d^4x \bar{E}_{p'}(x) E_p(x)~ && = (2 \pi)^4 
\hat{\delta}^{(4)}(p-p')
\nonumber \\ ~ && \equiv (2 \pi)^4 \delta_{kk'} 
\delta(p_0 - p'_0) \delta(p_2 - p'_2) \delta(p_3 - p'_3), 
\label{ortho}
\end{eqnarray}
\begin{equation} 
\Sigma \!\!\!\!\! \displaystyle{\int} d^4p E_p(x) \bar{E}_p(y)
\equiv \sum_{k} \int dp_0 dp_2 dp_3 E_p(x) \bar{E}_p(y)  
= (2 \pi)^4 \delta^{(4)}(x-y),
\label{complete}
\end{equation}
provided that the normalization constant in Eq.(\ref{eigenfcn}) 
is taken to be $N(n) = (4 \pi |gH|)^{1/4}/\sqrt{n!}$.  They also 
satisfy the useful relation\cite{Ritus}: 
\begin{equation}
\gamma \cdot \Pi~E_p(x) = E_p(x)~\gamma \cdot \bar{p},
\label{PROPERTY}
\end{equation}
where $\bar{p}_0 = p_0,~\bar{p}_1 = 0,~\bar{p}_2 
= - {\rm sgn}(gH) \sqrt{2|gH|k},~\bar{p}_3 = p_3$.  Note that, 
in terms of the momentum $\bar{p}$, the system is effectively a 
(2+1)-dimensional one.

These properties of the $E_p$-functions enable us to 
introduce the $E_p$-representation of the fermion 
Green's function: 
\begin{eqnarray}
G_A(p,p') &~\equiv~& \int d^4x d^4y \bar{E}_p(x) G_A(x,y) 
E_{p'}(y) \nonumber\\ &~=~& (2 \pi)^4 
\hat{\delta}^{(4)}(p-p') \frac{1}{\gamma 
\cdot \bar{p} + \tilde{\Sigma}_A(\bar{p})}
\label{pGreenfcn}
\end{eqnarray}
where $\tilde{\Sigma}_A(\bar{p})$ represents the eigenvalue 
matrix of the mass operator:
\begin{equation}
\int d^4x' M(x,x') E_p(x') = E_p(x) \tilde{\Sigma}_A(\bar{p}).
\label{masseigeneq}
\end{equation}
It is straightforward to verify that the inverse transform, 
\begin{equation}
G_A(x,y) = \Sigma \!\!\!\!\!\! \int \frac{d^4p}{(2 \pi)^4} E_p(x) 
\frac{1}{\gamma \cdot \bar{p} + \tilde{\Sigma}_A(\bar{p})} 
\bar{E}_p(y),
\label{Greenfcn}
\end{equation}
satisfies Eq.(\ref{Greeneq}).  Eqs.(\ref{pGreenfcn}) and 
(\ref{Greenfcn}) are the generalization of the well-known 
relations between coordinate space and momentum space Green's 
functions, with the plane wave eigenfunctions in the Fourier 
transform replaced here by the $E_p$-eigenfunctions in order 
to account for the presence of the external field.  
Eq.(\ref{pGreenfcn}) shows explicitly that the fermion 
propagator is diagonal (in momentum) in the 
$E_p$-representation.  As stated earlier, the mass operator 
is also diagonal in this representation: 
\begin{eqnarray}
M(p,p') ~&&~= \int d^4x d^4x' \bar{E}_p(x) M(x,x') E_{p'}(x')
\nonumber\\ ~&&~=~(2 \pi)^4 
\hat{\delta}^{(4)}(p-p') \tilde{\Sigma}_A(\bar{p}).
\label{pmassop}
\end{eqnarray}

Transforming to the $E_p$-representation, the SD equation, 
Eq.(\ref{SDx}), becomes\cite{LLN}
\begin{eqnarray}
\tilde{\Sigma}_A(\bar{p}) \delta_{kk'} 
&~=~&
i g^2 \sum_{k''} \sum_{\{\sigma\}} \int \frac{d^4q}{(2 \pi)^4}~ 
\frac{{\rm e}^{i {\rm sgn}(gH)(n-n''+\tilde{n}''-n')\varphi}}
{\sqrt{n!n'!n''!\tilde{n}''!}}~ 
\frac{{\rm e}^{- \hat{q}_\perp^2}}{q^2} \nonumber \\
& &
\cdot~ J_{nn''}(\hat{q}_\perp) J_{\tilde{n}''n'}(\hat{q}_\perp) 
\cdot \Delta \gamma^\mu \Delta'' \frac{1}{\gamma \cdot 
\bar{p}'' + \tilde{\Sigma}_A(\bar{p}'')} \tilde{\Delta}'' 
\gamma_\mu \Delta',
\label{SDsimp}
\end{eqnarray}
where
\begin{equation}
J_{nn''}(\hat{q}_\perp) \equiv \sum_{m=0}^{{\rm min}(n,n'')} 
\frac{n! n''!}{m! (n-m)! (n''-m)!} [i {\rm sgn}(gH) 
\hat{q}_\perp]^{n+n''-2m},
\label{defJ}
\end{equation}
\begin{equation}
\hat{q}_\perp^2 \equiv \frac{q_1^2 + q_2^2}{2|gH|}, ~~~~~~
\varphi \equiv \arctan \left(\frac{q_2}{q_1}\right),
\end{equation}
and the momentum $\bar{p}''$ is given by: 
$\bar{p}''_0 = p_0 - q_0$, $\bar{p}''_1 = 0$, 
$\bar{p}''_2 = -~{\rm sgn}(gH) \sqrt{2|gH|k''}$, 
$\bar{p}''_3 = p_3 - q_3$.  Here $n' = n(k',\sigma')$, 
$n'' = n(k'',\sigma'')$, $\tilde{n}'' = 
n(k'',\tilde{\sigma}'')$, $\Delta' = \Delta(\sigma')$, 
$\Delta'' = \Delta(\sigma'')$, $\tilde{\Delta}'' = 
\Delta(\tilde{\sigma}'')$, and the summation over 
$\{\sigma\}$ means summing over $\sigma$, $\sigma'$, 
$\sigma''$, and $\tilde{\sigma}''$.  Eq.(\ref{SDsimp}) 
is valid in the Feynman gauge.  The issue of gauge 
dependence has been addressed recently in Ref.\cite{FdlI} 
which shows that, within the approximations used, the 
solution\cite{LNA,LLN} to this equation which will be 
discussed below satisfies the Ward-Takahashi identities.

A general analytic solution to Eq.(\ref{SDsimp}) for 
$\tilde{\Sigma}_A(\bar{p})$ is not yet available.  
However, one can obtain an approximate infrared solution 
by the following simplifications.  First we observe 
that, due to the factor e$^{- \hat{q}_\perp^2}$ in the 
integrand, only the contributions from small values of 
$\hat{q}_\perp$ are important.  We may therefore 
truncate the $J_{nn''}$ series and keep only the terms 
with the smallest power of $\hat{q}_\perp$, i.e., 
$J_{nn''}(\hat{q}_\perp) \rightarrow n!~\delta_{nn''}$; 
and similarly for the $J_{\tilde{n}''n'}(\hat{q}_\perp)$. 
This will be referred to as the small $\hat{q}_\perp$ 
approximation and is valid only for weak couplings 
($g^2/4 \pi \ll 1$)\cite{LLN}.

Next, we sum over the spin indices and note that the 
remaining summation over $k''$ involves at most three 
terms: for $k > 0$, $k'' = k,~k \pm 1$.  In the limit 
$k$ = 0 = $\bar{p}_\perp$, we keep only the dominant 
$k'' = 0$ term.  This is known as the lowest Landau 
level approximation\cite{GMS}.  The SD equation is 
now simplified to 
\begin{equation}
\Sigma_A(\bar{p}_\parallel) ~\simeq~
2 g^2 \int \frac{d^4 q}{(2 \pi)^4} 
\frac{{\rm e}^{- \hat{q}_\perp^2}}{q^2} 
\frac{\Sigma_A(\bar{p}_\parallel-q_\parallel)}
{(\bar{p}_\parallel - q_\parallel)^2 
+ \Sigma_A^2(\bar{p}_\parallel-q_\parallel)}, 
\label{SDsigma}
\end{equation}
where we have made a Wick rotation to Euclidean space: 
$p_0 \rightarrow ip_4$, $q_0 \rightarrow iq_4$.  Note 
that the fermion wavefunction renormalization vanishes 
in the Feynman gauge, hence the self-energy 
$\tilde{\Sigma}_A$ has been replaced by the dynamically 
generated fermion mass $\Sigma_A$ in Eq.(\ref{SDsigma}).  
Except for the exponential factor in the integrand, 
Eq.(\ref{SDsigma}) has the same form as the corresponding 
SD equation when the external field is absent.  The 
difference is that only the longitudinal momentum is 
relevant here.  This reduction of dimensions from 4 to 
2 has been stressed in Ref.\cite{GMS}.

Finally, we consider the $\bar{p}_\parallel = 0$ limit 
and approximate the $\Sigma_A(q_\parallel)$ in the 
resultant integrand by $\Sigma_A(0) \equiv m$ to secure 
the gap equation,  
\begin{eqnarray}
1 &~\simeq~& 2 g^2 \int \frac{d^4 q}{(2 \pi)^4} 
\frac{{\rm e}^{- \hat{q}_\perp^2}}{q^2} 
\frac{1}{q_\parallel^2 + m^2} \nonumber \\
&~\simeq~& \frac{g^2}{4 \pi^2} |gH| \int_0^\infty 
d\hat{q}_\perp^2 \frac{{\rm e}^{- \hat{q}_\perp^2} 
\ln(2|gH| \hat{q}_\perp^2/m^2)}{2|gH| \hat{q}_\perp^2 - m^2}. 
\label{gapeq}
\end{eqnarray}
The solution to Eq.(\ref{gapeq}) has the form 
\begin{equation}
m \simeq a~\sqrt{|gH|}~{\rm e}^{- 2 \pi b/g},
\label{dynamass}
\end{equation} 
where $a$ and $b$ are positive constants of order 1.  The  
$g$-dependence of $m$ clearly indicates the nonperturbative 
nature of this result.

The above solution for the fermion dynamical mass is consistent 
with that found by Gusynin {\it et al.}\cite{GMS}, who studied 
the Bethe-Salpeter equation for the bound-state Nambu-Goldstone 
bosons of the spontaneously broken chiral symmetry.  The order 
parameter of this symmetry breakdown is computed to 
be\cite{LLN,SS}
\begin{eqnarray}
\langle \bar{\psi} \psi \rangle 
&~\simeq~& 
-~\frac{|gH|}{2 \pi^2} ~m ~\ln\left(\frac{|gH|}{m^2}\right)
\nonumber \\
&~\simeq~&
-~\frac{2ab}{g \pi} ~|gH|^{3/2} ~{\rm e}^{- 2 \pi b/g}.  
\end{eqnarray}
Note that, by allowing the fermion field $\psi$ to carry  
quantum numbers of internal symmetries, $\langle \bar{\psi} 
\psi \rangle \rightarrow \langle \bar{\psi}_i \psi_j \rangle$, 
the formalism developed above may be applied to study the 
dynamical breaking of internal gauge symmetries (in quenched, 
ladder approximation) in the presence of an external field.

Currently there is no experimental evidence for the magnetic 
field induced dynamical chiral/gauge symmetry breaking found 
in the above solution.  However, suggestive hints are 
available in excitonic systems.  In their recent experiment 
with the coupled AlAs/GaAs quantum wells\cite{exciton}, 
Butov {\it et al.}, found evidence of exciton condensation 
(condensation of electron-hole pairs analogous to the 
fermion-antifermion pairing in the spontaneously broken 
chiral vacuum) in the form of a huge broad band noise in the 
photoluminescence intensity when a sufficiently strong 
magnetic field is applied, while no exciton condensation 
occurs in the absence of the external magnetic field.

As an application, the method developed in the study of 
chiral symmetry breaking in an external magnetic field has 
been employed in the study of the effects of external 
magnetic fields on high-$T_c$ superconductors\cite{FM}.  
The dynamical chiral symmetry breaking solution found above 
may also be relevant for the chiral phase transition in 
heavy-ion collisions or in the electroweak phase transition
\cite{GMS} in the early universe when a large primordial 
magnetic field is expected to be present\cite{HEW}.  In 
these cases, the effects of temperature and chemical 
potential need be incorporated.

It is straightforward to generalize the above formalism to 
include nonzero temperature ($T \ne 0$) and nonzero chemical 
potential ($\mu \ne 0$) effects.  One finds that the gap 
equation, Eq.(\ref{gapeq}), becomes\cite{LLN}
\begin{eqnarray}
 1 &~\simeq~& \frac{g^2}{8 \pi^2} |gH| \int_{-\infty}^\infty 
 dq_3 \int_0^\infty d\hat{q}_\perp^2~  
 \frac{{\rm e}^{- \hat{q}_\perp^2}}{Q_1 Q_2}
\nonumber \\
 & & \cdot \left\{ Q_1 \left[ \frac{\coth(\frac{Q_2}{2T})}
 {Q_1^2 - (Q_2+\mu-i\pi T)^2} + \frac{\coth(\frac{Q_2}{2T})}
 {Q_1^2 - (Q_2 - \mu + i \pi T)^2} \right] \right. 
\nonumber \\
 & & ~~~~+ \left. Q_2 \left[ \frac
 {\tanh( \frac{Q_1+\mu}{2T})}{Q_2^2-(Q_1 + \mu - i \pi T)^2} 
 + \frac{\tanh(\frac{Q_1-\mu}{2T})}
 {Q_2^2-(Q_1 - \mu + i \pi T)^2} \right] \right\}_, 
\label{gaptu}
\end{eqnarray}
where  $~Q_1^2 \equiv q_3^2 + m^2_{T \mu}$, $~Q_2^2 
\equiv q_3^2 + 2 |gH| \hat{q}_\perp^2$, and $m_{T \mu}$ is 
the infrared dynamical fermion mass which depends on both 
the temperature and the chemical potential.  We use units 
in which the Boltzmann constant equals 1.  Aside from the 
quenched, ladder approximation, and the small 
$\hat{q}_\perp$ and lowest Landau level approximations, 
Eq.(\ref{gaptu}) is exact in its dependence on the 
coupling constant, the magnetic field, the temperature, 
and the chemical potential.

We have considered separately the $\mu = 0$ and $T = 0$ 
limits of Eq.(\ref{gaptu}).  We examine both analytically 
and numerically the behavior of the dynamical mass as 
$T$ (or $\mu$) is varied.  In the ($T \ne 0$, $\mu = 0$) 
case, we find that $m_{T 0}$ decreases monotonically as 
the temperature is raised and eventually vanishes above 
a critical temperature, indicating that the chiral 
symmetry is restored at high temperatures.  The critical 
temperature at which this continuous phase transition 
takes place is estimated to be\cite{LLN,GS,EZ}
\begin{equation}
T_c \sim 2 \sqrt{2}~m_{0 0},
\label{Tc}
\end{equation}
where $m_{0 0}$ is the ($T = 0$, $\mu = 0$) solution 
found in Eq.(\ref{dynamass}).  The order parameter for 
this phase transition exhibits similar behaviors:
\begin{eqnarray}
 \langle \bar{\psi} \psi \rangle_{T0} 
 &~\simeq~& 
 -~\frac{\vert gH \vert}{2 \pi^2} m_{T 0} \ln \left(
 \frac{\vert gH \vert}{m_{T 0}^2} \right) 
\nonumber \\
 &~\sim~&
 \vert gH \vert^{3/2} \left(1 - \frac{T}{T_c} \right)^{1/2} 
 \ln \left(1 - \frac{T}{T_c} \right)^{1/2},~~~~
 {\rm as} ~ T \rightarrow T_c^-.
\label{ccT}
\end{eqnarray}
The last expression reflects the behavior of $m_{T 0}$ 
near $T_c$:
\begin{equation}
 m_{T 0} \sim \vert gH \vert^{1/2} \left(1 - \frac{T}{T_c}
 \right)^{1/2},~~~~{\rm as} ~ T \rightarrow T_c^-.
\label{mc}
\end{equation}

In the ($T = 0$, $\mu \ne 0$) case, $m_{0 \mu}$ also 
vanishes as $\mu$ is increased beyond a critical value, 
thus restoring the chiral symmetry at high chemical 
potentials.  However, this chiral phase transition is 
discontinuous\cite{LLN}:
\[ \langle \bar{\psi} \psi \rangle_{0 \mu} ~\simeq~ 
 -~\frac{\vert gH \vert}{2 \pi^2} m_{0 \mu} \ln \left(
 \frac{\vert gH \vert}{m_{0 \mu}^2} \right)_, \]
\begin{eqnarray}
 m_{0 \mu} &~>~& 0,~~~~~~\mu < \mu_c, \nonumber \\
 &~=~& 0,~~~~~~\mu > \mu_c.
\end{eqnarray}
The critical chemical potential is approximately
\begin{equation}
 {\mu}_c \simeq \frac{m_{0 0}}{\sqrt{1 - \frac{2 I_2}
 {I_1}}}
\label{mucrit}
\end{equation}
where 
\begin{eqnarray}
 I_1 &~=~& \int_{-\infty}^{\infty} dq_3 \int_{0}^{\infty} 
  d\hat{q}_\perp^2~ \frac{{\rm e}^{-\hat{q}_\perp^2}}{Q^2 
  Q_2 (Q + Q_2)} \left(\frac{1}{Q} + \frac{1}{Q + Q_2} 
  \right)_, \nonumber \\
 I_2 &~=~& \int_{-\infty}^{\infty} dq_3 \int_{0}^{\infty} 
  d\hat{q}_\perp^2 ~\frac{{\rm e}^{-\hat{q}_\perp^2}} 
  {Q Q_2 (Q + Q_2)^3}_,   
\end{eqnarray}
and $~Q^2 \equiv q_3^2 + m_{0 0}^2$.  The integrals $I_1$ 
and $I_2$ are both positive and finite.

With these results, we can evaluate whether the dynamics 
of chiral symmetry breaking in a magnetic field discussed 
here may be relevant for the electroweak phase transition 
in the early universe.  For instance, the electroweak phase 
transition took place at a temperature of order 100 GeV.  
From Eq.(\ref{Tc}), this requires a magnetic field of 
order $2 \times 10^{41}$ gauss if we take $a = b = 1$ and 
$4 \pi/g^2 \simeq 137$.  This is much larger than any 
estimates of the magnetic field strength at the time of 
the electroweak phase transition\cite{HEW}.  We may 
therefore conclude that the chiral symmetry breaking 
solution considered here does not play any role in the 
electroweak phase transition.

On the other hand, as suggested earlier, the formalism 
described here can be used to study the dynamical breaking 
of internal gauge symmetries, abelian as well as nonabelian, 
in the presence of a magnetic field.  Within the quenched, 
planar approximation, we expect the same generic results 
as obtained here to be applicable in those situations.  
We may therefore entertain the possibility that the 
coupling constant is relatively large, e.g., $4 \pi/g^2$ 
of order 0.1 (As concrete examples, we may consider 
electroweak symmetry breaking in technicolor models or 
chiral symmetry breaking in QCD by colored fermions 
belonging to large representations of SU(3)).  In this case 
the required magnetic field could be of order $10^{28}$ 
gauss or less, which makes it an interesting possibility 
for the study of phase transitions in the early universe.

\vspace*{1.0cm}
\begin{center} 
{\bf ACKNOWLEDGEMENTS}\\
\end{center}

This work was supported in part by the U.S. Department of Energy 
under Grant No. DE-FG02-84ER40163.  I would like to thank the 
faculties at the Physics Departments of the Tokyo Metropolitan 
University and of the Academia Sinica in Taipei, especially 
H. Minakata, O. Yasuda, and H.-L. Yu, for their hospitality 
during my visit to their institutions where this article was 
written.
\raggedbottom

\end{document}